\documentclass[12pt,preprint]{aastex}
\slugcomment{Accepted and scheduled for publication  
in {\it the Astrophysical Journal},  for the ApJ 20 June 2005, v 626 2 issue}  
\def\lax {\ifmmode{_<\atop^{\sim}}\else{${_<\atop^{\sim}}$}\fi}  
\def\gax {\ifmmode{_>\atop^{\sim}}\else{${_>\atop^{\sim}}$}\fi}  
\def\gtorder{\mathrel{\raise.3ex\hbox{$>$}\mkern-14mu
             \lower0.6ex\hbox{$\sim$}}}
\def\etal { et al. }

\begin{document}

\title{How To Distinguish Neutron Star and 
Black Hole X-ray Binaries? Spectral index and Quasi-Periodic Oscillation Frequency  
Correlation}

\author{Lev Titarchuk\altaffilmark{1,2} and Nickolai Shaposhnikov\altaffilmark{3}}

%\altaffiltext{1}{University of Maryland, College
%Park:rfiorito@UMD.edu}
\altaffiltext{1}{George Mason University/Center for Earth
Observing and Space Research, Fairfax, VA 22030; and US Naval Research
Laboratory, Code 7655, Washington, DC 20375-5352; ltitarchuk@ssd5.nrl.navy.mil }
\altaffiltext{2}{Goddard Space Flight Center, NASA, Exploration of the Universe Division, code 661, Greenbelt  
MD 20771; lev@milkyway.gsfc.nasa.gov}
\altaffiltext{3}{Goddard Space Flight Center, NASA, Exploration of the Universe Division/Universities Space Research Association, code 662, Greenbelt  
MD 20771; nikolai@milkyway.gsfc.nasa.gov}
%  rfiorito@milkyway.gsfc.nasa.gov; lev@lheapop.gsfc.nasa.gov}

\begin{abstract}
Recent studies have revealed strong correlations between  1-10 Hz frequencies
of quasiperiodic oscillations (QPOs)    and the
spectral power law index of several Black Hole (BH) candidate sources when seen
in the low/hard state, the steep power-law (soft) state, and in transition 
between these states. In the soft state these index-QPO frequency 
correlations show a saturation  of the photon index  
$\Gamma\sim 2.7$ at high values of the low frequency $\nu_{L}$. 
This saturation effect was previously identified as a black hole signature.   
In this paper we argue that this saturation does not occur, at least 
for one neutron star (NS) source 4U 1728-34, for which the index 
$\Gamma$ monotonically 
increases with $\nu_{L}$ to the values of 6 and
higher. We base this conclusion on our analysis of $\sim 1.5$ Msec 
of RXTE archival data for 4U 1728-34. 
We reveal the spectral evolution of the Comptonized blackbody spectra  
when the source transitions from the hard to soft states. 
 The hard state spectrum is a typical  thermal
Comptonization spectrum of the soft photons  which originate in the 
disk and the NS outer photospheric layers.
 The hard state  photon index is $\Gamma\sim 2$. The soft state  spectrum  
 consists of two blackbody components which are only slightly Comptonized. 
%(the Comptonization Green's function index, $\Gamma>6$).  
%The color disk and NS surface temperatures  slightly decreases
%from 0.93 keV to 0.83 keV and from  2.9 keV to 2.2 keV respectively when the source undergoes  
%the transition from the hard to soft states.
Thus we can claim (as expected from theory) that in NS sources  thermal
 equilibrium is established for the soft state.
To the contrary in BH sources, 
the equilibrium is never established due to the presence of the BH horizon.
The emergent BH spectrum, even in the high/soft state, has 
a power law component.  
We also identify the low QPO frequency $\nu_L$ 
as a fundamental frequency of the 
quasi-spherical component of the transition layer (presumably related
 to the corona and the NS and disk magnetic closed  field lines).  
%between the surface NS and the Keplerian disk
The lower frequency $\nu_{SL}$ is identified as the frequency of oscillations of
a quasi-cylindrical configuration of the TL  
(presumably related to the NS and  disk magnetic open field lines).
We also show that the presence of Fe K$_{\alpha}$ emission-line strengths, QPOs, and the link between them
does not depend on radio flux  in 4U 1728-34.

\end{abstract}

\section{Introduction}

One of the  basic questions addressed in many observational and theoretical 
studies is how can one  observationaly distinguish a neutron star from a black hole.
Before the {\it Rossi X-ray Timing Explorer (RXTE)}
era most of the efforts to identify black hole (BH) and neutron star (NS)  
were primarily concentrated on the observations of features  in the energy spectra of these objects.
However, one attempt to use timing behavior to identify the nature of
compact object can be mentioned. Specifically, 
QPO phenomena discovered in some of the bright 
low mass X-ray binaries (LMXBs) in the 6-60 Hz range by EXOSAT  (see e.g. van 
der Klis et al. 1985) were considered to be NS signatures and were
interpreted as that time as a beat effect between  the inner edge of an 
accretion disk and a neutron star magnetosphere (Alpar \& Shaham 1985, see also Lamb et al. 1985).

The first attempts to look for the effect of the presence or absence of a
surface on the spectrum concentrated on comparing BH and NS stars in the 
high/soft state when the mass accretion rate
in the source is presumably high. Chakrabarti \& Titarchuk (1995), hereafter 
CT95,  Titarchuk, Mastihiadis \& Kylafis (1996), (1997), hereafter TMK96, 
TMK97 respectively,  Titarchuk \& Zannias (1998), hereafter TZ98, and 
Laurent \& Titarchuk (1999, 2001), hereafter LT99, LT01 respectively, argue 
that in BHs the specific extended power-law spectrum is formed with a photon 
index of 2.5-2.7 (depending on the plasma temperature of the accretion flow) 
due to the bulk motion Comptonization in the converging flow into the BH. On the other 
hand, in a high mass accretion rate NS, 
%the soft photons emitted 
%by the NS surface are completely comptonized and 
the emergent spectrum acquires  blackbody like shape (see LT99 and TMK96).

Narayan, Garcia \& McClintock (1997, 2002) and \citet{Garcia} introduced
the idea of comparing the NS and BH in quiescence. They suggested that the 
 advection dominated flow (ADAF) into a black 
hole is radiatively inefficient. Hence, the NS boundary layer should
be more luminous with  respect to BH for  the same mass accretion rate.
However, while NSs are somewhat more luminous than BH systems, the difference is 
not nearly as large as predicted by the ADAF models \citep{men}.  

Titarchuk \& Zannias (1998) calculated the space distribution of the high 
energy photons produced in the converging flow by the bulk motion 
Comptonization. Later LT01  confirmed that this distribution
peaks between 2 and 3 Schwarzchild radii, $R_{\rm S}$.  
This result led them to the prediction that any high frequency oscillation 
occurring in the innermost part of the accretion disk should be seen as
a quasi-periodic oscillation (QPO)  of the high energy emission from  
converging flow (the extended power law) but not as a QPO of the soft blackbody 
emission of the accretion disk. This is to be expected because the 
effective area of the soft X-ray disk emission, $\sim \pi (15 R_{\rm S})^2$ is at least of 
one order of magnitude larger than that of the innermost part of the disk, 
$\sim \pi (3R_{\rm S})^2$. Thus the QPOs of 100-200 Hz frequencies are 
seen as the oscillations of the extended steep power-law component,
 and should be considered to be a BH signature. This consequence 
of the Bulk motion Comptonization is supported by the observational 
correlation of the extended power law with the high frequency QPOs 
(see details of the observations in a review by McClintock \& Remillard 2003).

\citet{SR00} and \citet{BPK} suggested that one could discriminate between 
NS and BH systems using their timing signatures.
Particularly, \citet{SR00} analyzed 
%the observational evidence analyzing PDSs of 
a sample of 9 NS and 9 BH in the low/hard
spectral states and presented  observational evidence 
that  in the PDS of accreting neutron stars with weak magnetic field
significant power is contained at frequencies close to one kHz.
For most Galactic black holes, the power spectrum demonstrates 
a strong decline at frequencies higher than 10-50 Hz.
Using the data analysis of power spectra of NS and BH sources \citet{BPK}
found
 % implies that the difference in PDSs of BHs and NSs is mostly 
%caused by scaling effect, namely 
that all characteristic frequencies are lower in the case of the BH, probably reflecting the mass 
dependence of the dynamical timescale in accretion flow 
(see more evidence of this mass scaling of QPO frequencies in Titarchuk \& 
Fiorito 2004, hereafter TF04). Another important 
effect should be noted. \citet{hom2003} found that the power of the high-frequency QPOs
($\sim 200$ Hz) is comparable and even sometimes higher than that of 
the low-frequency QPOs ($\sim 10$ Hz) during the transition from low/hard to 
high/soft states of BH sources. For this transitional state they  reported 
the detection of the QPO high frequency at 250 Hz in 
black hole transient XTE J1650-500 and also found that its strength 
changed with the total count rate in the 2.5-60 keV band. The QPO power 
also increased with energy with rms amplitudes of $<0.85$\%, $<4.5$\% and $<12.1$
\% respectively in 2-6.2, 6.2-15.0 and 15.0-60 keV bands (see Fig. 2 in Homan 
et al. 2003).

The well-known absence of type I X-ray bursts in BH candidate sources leads 
Narayan \& Heyl (2002) to conclude that these sources have event horizons.  
They have carried out a linear stability analysis of the accumulating fuel 
on the surface of a compact star to identify the conditions
under which thermonuclear bursts are triggered. It is also shown that, if BH 
candidates had surfaces, they would very likely exhibit
instabilities similar to those that lead to type I bursts on neutron stars.

In  early spectral studies of BHs, and  the identification of the BH
signatures by Shrader \& Titarchuk (1998), (1999), (2003), 
Borozdin et al. (1999), and Titarchuk \& Shrader (2002), 
the authors stated that,  in the high/soft states,  
 the $\sim 1$ keV thermal emission along with the steep power-law 
tail of the photon index 2.5-2.7 are the observational evidence of the BH signature.
This  particular spectral BH signature was predicted by CT95, TMK96, TMK97, 
TZ98, LT99, LT01 using the extensive  radiative transfer calculations of 
soft (disk) photon Comptonization in the advection dominated converging flows 
onto black hole and neutron stars. Furthermore, recently \citet{DG} compiled 
a large sample of high-quality data from {\it RXTE} Proportional Counter Array
 (PCA) observations of BHs and low-magnetic-field NSs (atolls and Z sources)
and showed that the source evolution as a function of luminosity is very 
different depending on the nature of the compact object. They also reported 
 a spectral state that is only seen in BH systems when the 
sources are bright. In this state the thermal emission, presumably, from a low temperature 
disc dominates, but also  a steep power-law tail to high energies is observed. 
They argue that this is a unique signature 
for any new transient to be identified as a black hole.
In fact, the Done \& Gierlinski's (2003) study supports our conclusions regarding 
% that this is  a unique black hole signature to be identified as a 
uniqueness of this black hole  signature.
%  The extensive spectral 
%data analysis and study of color-color diagram evolution 
%as a function of luminosity for NS and BH sources 
They come to 
% lead them
%\citet{DG} 
  the same conclusions  regarding {\it the spectral BH signature 
as the thermal emission along with the steep power-law tail  in 
the high/soft state of the source}.

In the context of drawing a distinction between NS and BH sources 
we would like to point out the recent paper by  \citet{torr} 
who offered a new method of identification of neutron star systems using 
a temperature-luminosity relation.    
The source can only be a NS if a soft blackbody-like component with the color temperature  
of order of  1 keV is observed in the spectrum of the low/hard state. 
On the contrary, such a high temperature is necessarily related to a high disk luminosity in the case  
of a BH and this case is never observed for BHs  in the low hard state. 
In fact, the low/hard state spectra are similar for 
BH and NS sources. Both of them are
fitted by Comptonization spectra with the index in the range of 1.5-1.8 and 
with the electron temperatures of the Comptonizing medium   in the range of 30-60 
keV. \citet{torr} applied this new method for identification
for a NS in the non-pulsating massive X-ray binary 4U2206+54.

However, one can argue that  similar spectra consisting of thermal emission 
and a steep power-law tail of the photon index 2.7 can be observed in NS 
sources. In fact, this type of spectra have been detected from Sco X-1 using X-ray 
observations from rockets in the very beginning of X-ray Astronomy 
(see Morrison 1967) and confirmed 30 years later by  Strickman \& Barret 
(1997) using GRO and RXTE observations of Sco X-1. It is worth noting that this photon index 2.7 is only one value which is necessarily indicative of saturation level observed in BHs (see Vignarca et al. 2003 and TF04).

Thus, to reveal the nature of a compact object, one has to study the evolution  
of the source X-ray spectra 
 as a function of luminosity (or mass accretion rate), but not just the
evolution of the color-color diagram [cf. Done \& Gierlinski (2003)].
  The actual difference in the BH and NS spectra should be expected in 
the high state when the thermal equilibrium
is established and a blackbody spectrum is formed in NS sources due to
 the presence of solid surface. 
This thermal equilibrium condition is never
satisfied for a BH because of the event horizon. 
In this paper we present a comprehensive spectral and timing analysis of
archival RXTE data from the NS source 4U 1728-34 in order to establish 
the spectral evolution of the source as a function of luminosity and its
timing characteristics of the power spectrum (PDS).
  
Recent studies by Vignarca et al. (2003) and TF04 have revealed strong correlations between 
low frequency QPOs (1-10 Hz) and the
spectral power law index of several Black Hole candidate sources when seen
in the low/hard state, the steep power-law (soft) state, or in transition 
between these states. The observations support the notion  that the presence of a power-law component in  
the low/hard state spectrum is related to simultaneous radio emission, which, in turn, indicates the probable presence of a jet 
[see e.g. \citet{mirabel,klein}].
%Strong QPOs ($>20\%$ rms) are present in the power density spectrum in the
%spectral range where the power-law component is dominant ( i.e. 60-90\% )
%This evidence contradicts the dominant long standing interpretation of QPOs as  
%a  signature of the thermal accretion disk.

TF04 show that the observed low frequency QPO - spectral index correlation is a  
natural consequence of an adjustment of the Keplerian disk to the
innermost sub-Keplerian boundary conditions near the central object. This
ultimately leads to the formation of the sub-Keplerian transition layer (TL)
between the adjustment radius and the innermost boundary (the horizon for BH) [see details of the transition layer  
model (TLM) in Titarchuk, Lapidus \& Muslimov (1998), hereafter TLM98].
 In the framework of TLM  
% $\nu_{high}$ is related to the
%gravitational frequency at the outer (adjustment) radius $\nu_{g}\approx\nu_{\rm K}$ and
the low frequency $\nu_{L}$ is related to the magnetoacoustic oscillation frequency $\nu_{MA}$.  
Using a relation between $\nu_{MA}$, the mass accretion rate, the
photon index $\Gamma$ and the mass accretion rate TF04 infer a correlation 
between $\nu_{MA}$ and the spectral photon index $\Gamma$.
%(iii) identification of $\nu_{low}$ with $\nu_{MA}$, allow us to make a
%comparison of the theoretically predicted correlation with the observed
%correlation. For this identification we use the one temperature plasma
%assumption. We apply the plasma temperature obtained from the Comptonization
%spectra (electron temperatures) for calculations of magnetoacoustic
%frequencies which strongly depend on the proton temperature. The one
%temperature assumption is quite consistent with the data.
TF04 present strong arguments that in BHs the two particular distinct 
phases occur in which one of them, the steep power-law phase is the signature 
of a BH. They also found that a hard state (related to an extended 
Compton cloud) is characterized by the photon index at about 1.7 and the 
low QPO frequencies below 1 Hz. 
This is the regime where thermal Comptonization
dominates the upscattering of soft disk photons and the spectral shape (index)
is almost independent of mass accretion rate. TF04 find that the soft 
state is related to the very compact region
where soft photons of the disk are upscattered to form the steep power
law spectrum with photon index saturating at about 2.8. This is the regime where
Bulk Motion Comptonization dominates and the effect of an increase in the mass
accretion is offset by the effect of photon trapping in the converging flow
into the BH.

In this Paper we study the low frequency-index correlation for a NS source 
4U 1728-34. We find  that  this correlation is different  from that was previously found in BH sources (see TF04 and 
Vignarca et al. (2003) where  the low frequency-index correlation is presented  for large samples of BH sources).   
In  PDSs observed by the {\it RXTE} for 4U 1728-34,  
Ford \& van der Klis (1998, hereafter FV98) discovered low-frequency 
Lorentzian (LFL) oscillations  with frequencies $\nu_{L}$ between 10 and 50 Hz.
These frequencies as well as the break frequency $\nu_b$ 
were shown to be correlated with high frequency QPO peaks (FV98). 
We also note that nearly coherent (burst) oscillations with  
frequency $363\pm 2$ Hz was discovered by for Strohmayer et al. (1996) 
during the type I X-ray bursts in this source. \citet{to}
 presented a model for the radial 
oscillations and diffusion in the TL surrounding 
the neutron star.  
Titarchuk \& Osherovich, using dimensional analysis, have identified the corresponding radial oscillation  
and diffusion frequencies in the TL
 with the low-Lorentzian $\nu_{L}$ and break frequencies $\nu_b$ for 4U 1728-34.   
They predicted values for $\nu_b$ related to the diffusion in the boundary layer, that are consistent
 with the observed $\nu_b$. Both the Keplerian and radial oscillations, 
along with diffusion in the transition layer, are controlled  
 by the same parameter: the Reynolds number $\gamma$, which in turn is related to the accretion rate (see also TLM98).
It is worth noting that the nature of the break frequency as a diffusion effect (the inverse of  time
of the diffusion propagation in the bounded configuration) was later  corroborated  by Wood et al. (2001)
and Gilfanov \& Arefiev (2005). Particularly,  Wood et al. 
demonstrated that in BHC  XTE 1118+480  X-ray light curves with fast rise/exponential decay profile are a consequence 
of the diffusion matter propagation in the disk. On the other hand, Gilfanov \& Arefiev studied X-ray variability of
persistent LMXBs in the $\sim 10^{-8}-10^{-1}$ Hz frequency range aiming to detect features in their PDS associated with 
diffusion time scale of the accretion disk $t_{diff}$. As this is the longest intrinsic time scale of the disk, the power spectrum is
expected to be independent of the frequencies less than $\nu_b=1/t_{diff}$.  They found that the break frequency correlates very
well with the binary orbital frequency in a broad range of binary periods $P_{orb}\sim 12$ min - 33.5 days, in accord with
theoretical expectations for the diffusion time scale of the disk.
 
The observational signature of the mass accretion rate $\dot M$ is the QPO low frequency as has been shown for BH sources by TF04. 
The QPO frequency is not only related to  $\dot M$ but also the size of the Comptonizing region, $L$ i.e. $\nu_{QPO}\propto 1/L$.
The behavior of $\nu_{QPO}$ with respect to spectral index $\Gamma$ connects the characteristics of the Comptonization and spectral state with $\dot M$.
This is graphically represented for BHs in the observations of Vignarca et al. (2003). We will similarly employ this type of data to compare NS
spectral states with BHs to show their qualitative differences.

In \S 2 we present the details of the our spectral and timing  data analysis 
of archival RXTE data from the NS source 4U 1728-34.
In \S 3 we present and discuss the results of the data analysis and we compare them to that presented by TF04  for BH sources. 
In \S 4 we offer an explanation 
of the pair of QPO low  frequencies detected in  
4U 1728-34.  A link between Fe K$_{\alpha}$ emission-line strength and QPO frequencies in 4U 1728-34 is presented and 
disk-jet coupling in X-ray binaries is  discussed in \S 5. Conclusions  follow in \S 6.

\section{Observations and data analysis}

We analyzed the data collected by PCA (Jahoda et al. 1996),
the main instrument on  
board the {\it RXTE}. Observational {\it RXTE} data for 4U 1728-34 that are
available through the GSFC public archive \footnote{http://heasarc.gsfc.nasa.gov}. 
Data reduction and analysis was conducted with FTOOLS 5.3
software according to the recipes from ``{\it RXTE} Cook Book''.

The summary of the {\it RXTE} observation proposals and data used in the 
presented analysis is given in Table 1. Each proposal consists of a set
observations that can be divided into intervals of continuous on-source 
exposure (usually about 3 ks) corresponding to one {\it RXTE} orbit. For 
each proposal we provide  its archival identification number (proposal ID), 
the dates between which the data were collected, the total on-source exposure, the number of observations in the proposal $N_{obs}$,  the
number of continuous data intervals $N_{int}$,  the average number of operational PCUs in the proposal ${\bar{N}_{PCUon}}$ and a 
set of references related to the timing analysis of the persistent flux of
4U 1728-34. The data collected during the first two years of {\it RXTE} 
operation have been the main focus of the  data analysis and
interpretation (see Strohmayer et al. 1996; Ford \& van der Klis 1998; 
van Straaten et al. 2002; Di Salvo et al. 2001; Mendez, van der Klis \& Ford
2001; Jonker, Mendez \&  van der Klis 2000). Timing analysis of
4U 1728-43 data from years 1999, 2000 and 2001 was  presented in 
Migliari, van der Klis \& Fender (2003) where the authors 
studied correlation between the source  timing properties and its position 
on color-color diagram. 
The overall exposure of the data used is $\sim 1,500$ ks.

We calculated an energy spectrum and an averaged PDS for each continuous
interval of data (excluding X-ray bursts), which correspond to one 
orbital {\it RXTE} revolution. First we performed the data
screening to calculate good time intervals for Fourier analysis.
We excluded the data collected for elevation angles less than 
$20^\circ$ and during South Atlantic Anomaly passage.
To avoid the electron contamination we also applied 
the condition for electron rate in the PCU 2 (which is operational during
all observations) to be less than 0.1.
The data was rebinned  
to $2^{-12}$ second time resolution to obtain Nyquist frequency of
2048 Hz. Individual PDS were calculated for
each consecutive 8 second intervals, and then was averaged to
obtain a final PDS for the entire continuous interval. 
The PDS are normalized to give rms fractional variability per Hz.
Noise level is calculated between 1500 and 2000 Hz where no
source variability is expected and is subtracted before
applying any model fitting.
 For the PDS modeling we used the 
broken power law \citep[see][for definition]{VS00}  
component to fit broad band frequency noise  
and Lorentzians to describe QPO profiles.

We extract energy spectra from Standard2 data files using counts from
upper xenon layer of all operational detectors.
For the PDS calculation  we  used high resolution Event data mode which comprises counts
from all layers of all detectors which are on during a
particular observation. Spectral fitting was performed using the 
XSPEC astrophysical fitting package.

To describe  the shape of outgoing 
spectrum we use the {\it BMC} XSPEC model (Titarchuk et al. 1997) for both the NS and 
disk components. The BMC model spectrum  is the sum of
the (disk or NS)  black-body component and Comptonized black-body component.
It has four parameters: $kT$ is a  color temperature of thermal photon spectrum,
 $\alpha$ is the energy spectral index ($\alpha=\Gamma-1$,  where $\Gamma$ is the photon index),  
the parameter $A$  related to  the weight of the Comptonized component, $A/(1+A)$,
and a normalization of the blackbody component.

The radiation from central object and inner parts of accretion disk is 
Comptonized by a surrounding cloud.
As long as the spectral index  for each {\it BMC} model is a physical characteristic of the Comptonized region
(i.e. the Compton cloud), we tie these parameters together for the BMC models describing disk and 
NS spectral components.
%Thus all radiation from NS and inner disk passing through the cloud and get Comptonized.
We found during fitting procedure  that the relative weight of the Comptonized component $A/(1+A)$
is very close to one, in other words that {\it all} radiation from NS and inner disk passing through the cloud and get Comptonized.
Thus,  we freeze  {\it log A} parameter at 5.0 to ensure this equality $A/(1+A)=1$ is satisfied.
We also add a Gaussian of energy $\sim$ 6.4 keV, which is presumably
due to an iron emission line.
 We use a fixed hydrogen column of $N_H=1.6\times 10^{22}$ provided  
by HEASARC 
%\citep[][for details]{nH} 
[see Dickey \& Lockman (1990) for details] to model Galactic absorption.
Consequently  the XSPEC model for the spectral fitting reads as {\it WABS(GAUSSIAN+BMC+BMC)}
and we use $\alpha$, the NS and disk color temperatures, the NS and disk blackbody normalizations 
as free parameters of the spectral continuum model.  The spectral fits was obtained using 3.5-30.0 keV energy range.
We add 1\%  error to the data to account for systematic uncertainty in the  PCA 
calibration. The typical quality of fit is good with $\chi^2_{red}$ in the range of 0.3-0.9, and more than 90\% of
fits having  $\chi^2_{red}$ in the range 0.7-0.9. 
%Lower values of  $\chi^2_{red}$ are reached for 
%some short data intervals with exposure less 1000 seconds or for orbits where less than 3 detectors 
%were on as a result of insufficient counting statistics.

\section{Data analysis results. The spectral evolution and  index-frequency correlation and  in 4U 1728-34}
\begin{figure}[ptbptbptb]
\includegraphics[scale=0.65,angle=-90]{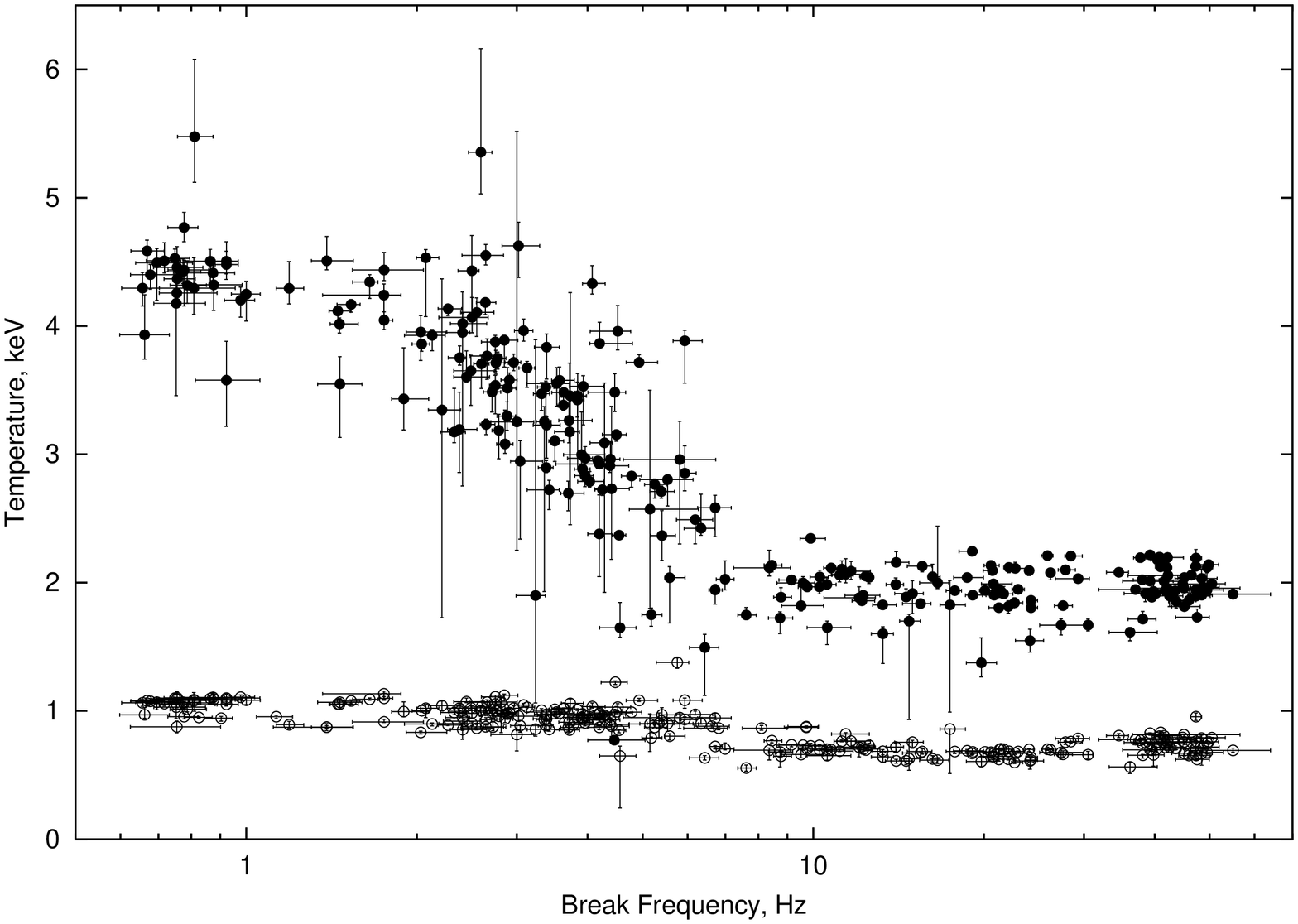}
\caption{
Temperatures of spectral components  for NS (filled circles) and accretion disk (empty circles).
Low/hard state corresponds to the left side of the picture, high/soft - to the right.
}
\label{kT}
\end{figure}

\begin{figure}[ptbptbptb]
\includegraphics[scale=1.2,angle=-90]{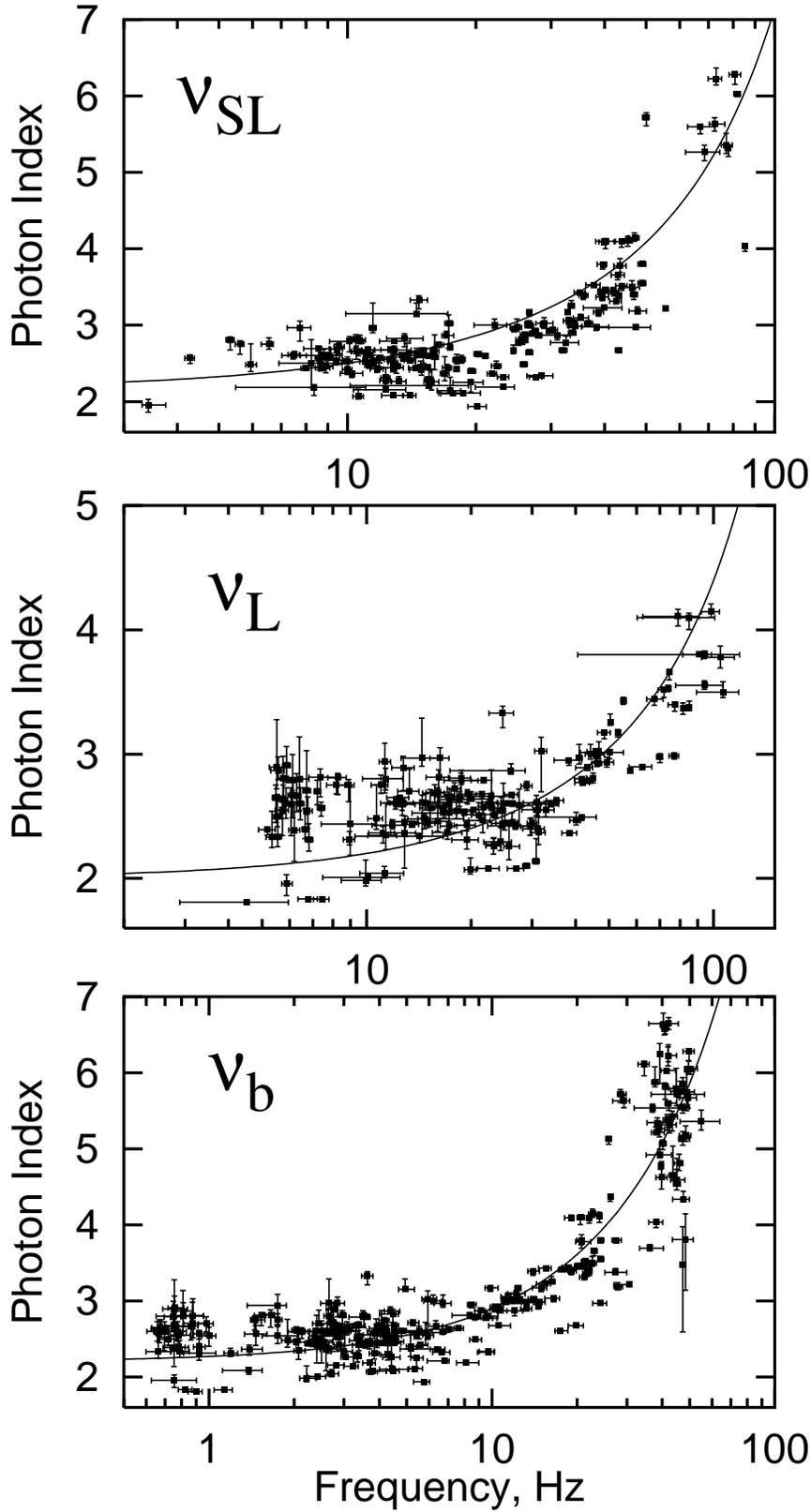}
\caption{
The observed  correlations
 between photon index and break frequency $\nu_b$ (lower panel), QPO low frequencies 
$\nu_L$ (middle panel), and  $\nu_{SL}$ (upper panel) for the NS source 4U 1728-34.}
\label{ind_qpo}
\end{figure}

%\subsection{The spectral evolution and  index-frequency correlation and  in 4U 1728-34}
The evolution of spectral properties of the source during the transition from low/hard to 
high/soft state is shown on Figures \ref{kT} and \ref{ind_qpo}.  The temperature profiles 
of two {\it BMC} spectral components are plotted versus break frequency $\nu_b$ in Figure 
\ref{kT}.  The higher temperatures presented by
filled circles are presumably related to the Comptonized NS blackbody component 
%orig the {\it color) temperature of the NS surface, while lower one is the temperature 
%of photons coming from accretion disk. 
which color temperature reaches $\sim 4.5$ keV  for the hardest states ($\nu_b \lesssim 1.0$ Hz). 
For $\nu_b$ between 2 and 8 Hz, it 
softens steeply to $\sim 2$ keV and levels off at this value for break frequencies higher than 10 Hz.
The soft (disk) color temperature changes only slightly, 
decreasing from $\sim 1$ keV in the low/hard state to  $\sim 0.8$ keV in the high/soft state. 

\begin{figure}[ptbptbptb]
\includegraphics[width=5.1 in,height=6.1in,angle=-90]{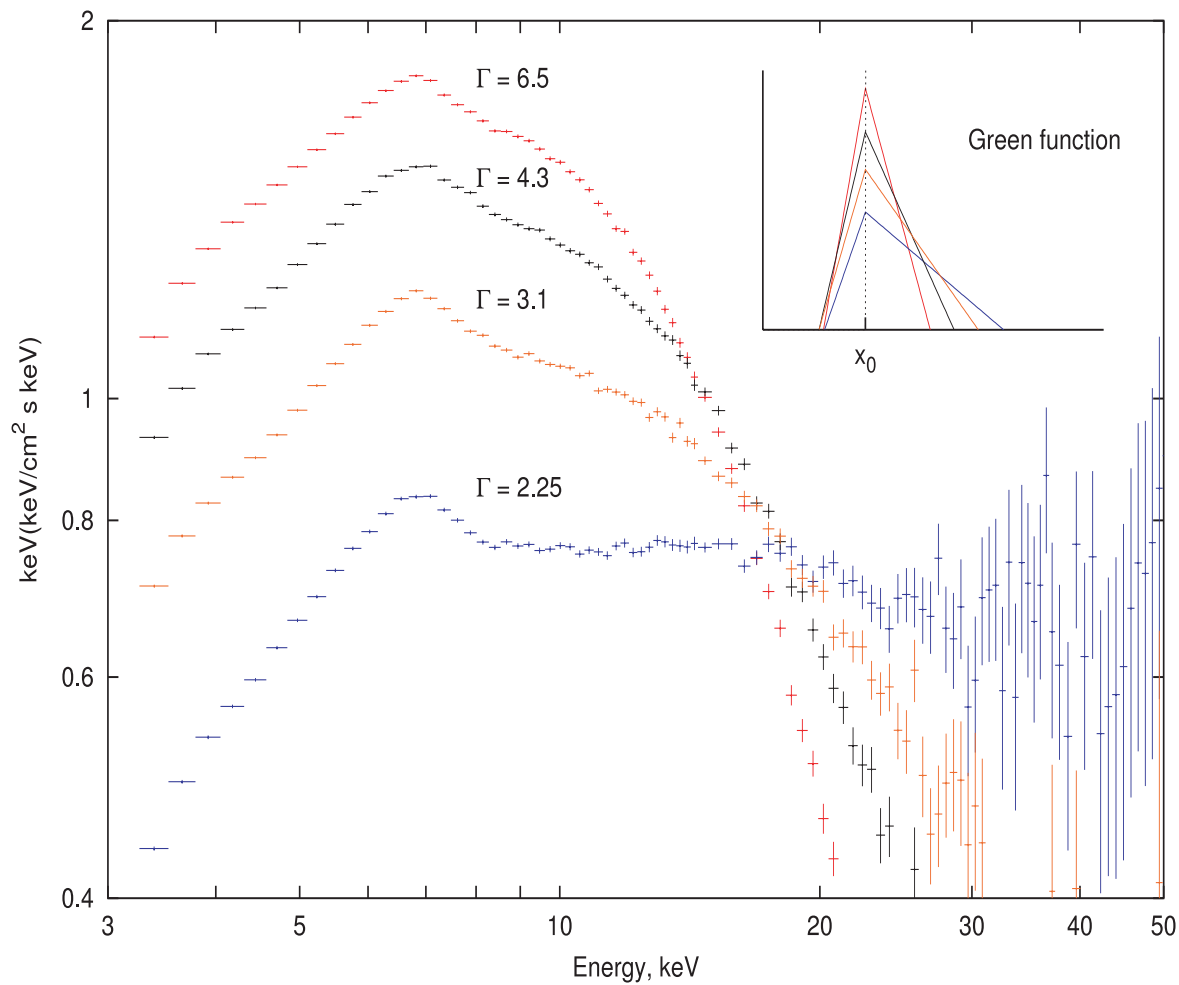}
\caption{
Spectral evolution of the source from  low/hard state to high/soft state. Photon index of the upscattering
Green function $\Gamma$ changes from 2.25 to 6.5 respectively.    
In the embedded panel we show the evolution of the upscattering Green function.  
One can clearly see an evolution of the broken power law with the high-energy power-law tail  
of index 2.25 to nearly a Delta function distribution.}
\label{evol}
\end{figure}

\begin{figure}[ptbptbptb]
\includegraphics[width=4.in,height=6.1in,angle=-90]{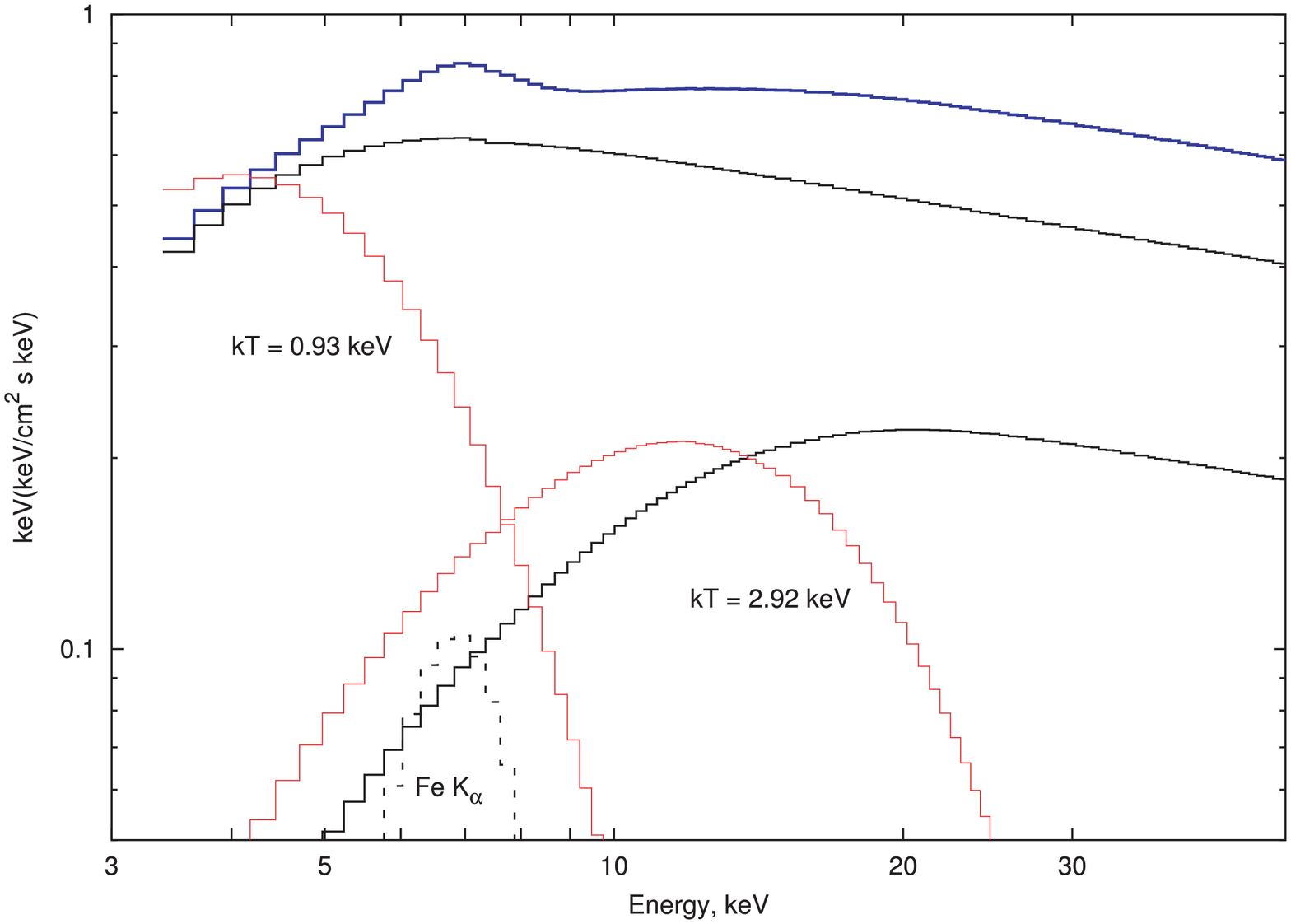}
\includegraphics[width=4.in,height=6.1in,angle=-90]{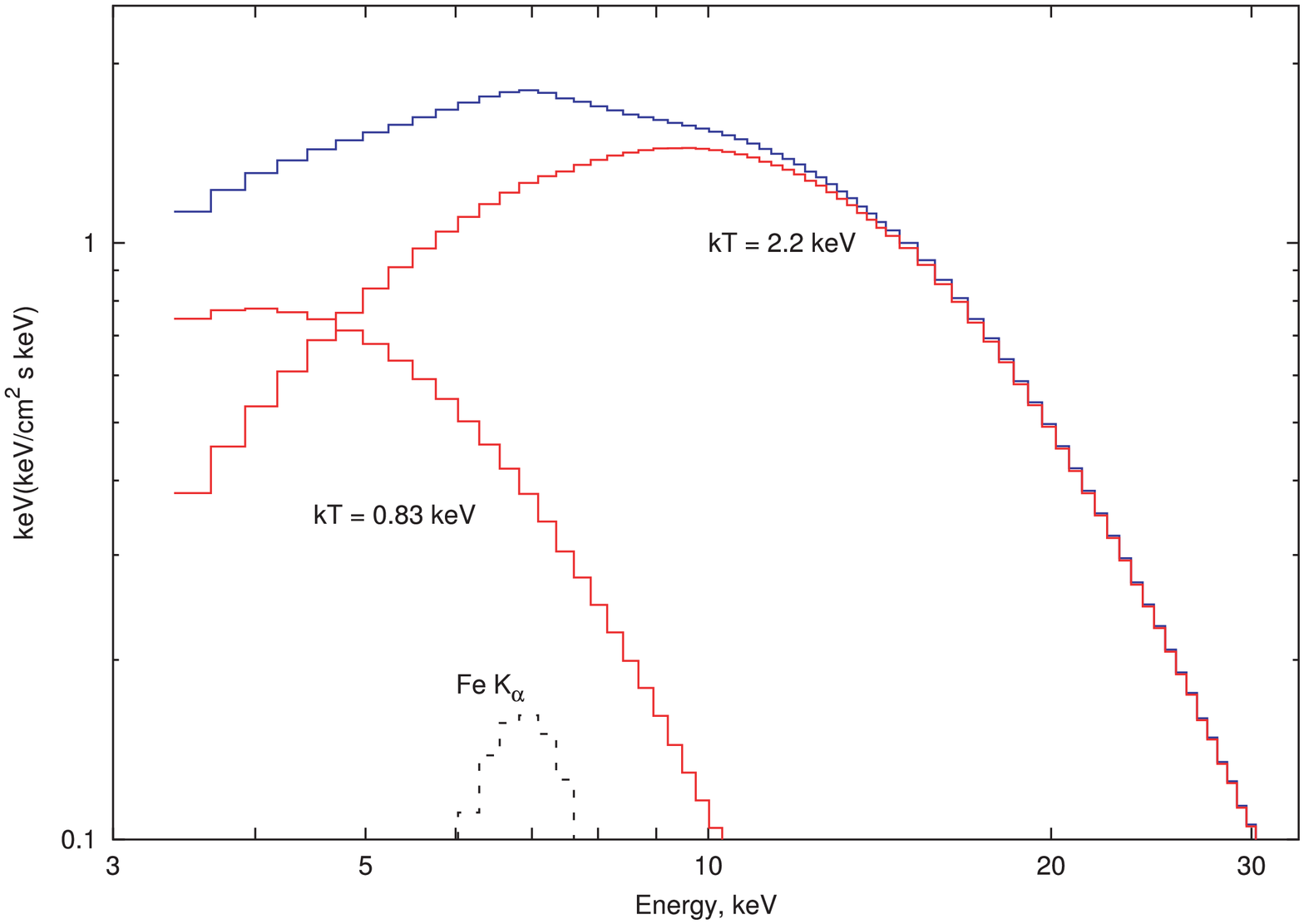}
\caption{
Spectral components  (blue curve) of  low/hard state (upper panel) and high/soft state 
(lower panel). The low/hard state spectrum consists of two Comptonized blackbody components 
(black curves). For  the soft blackbody components (red curves)  
 the color temperatures  are 0.93 keV and 2.92 keV and K$_{\alpha}-$iron line component (dashed curve). The high/soft state spectrum  
consists of two pure blackbody components (red)  for which color temperatures are    
0.83 keV and 2.2 keV and K$_{\alpha}-$iron line component.}
\label{soft_hard}
\end{figure}

In our timing study of the persistent radiation we reveal the following PDS features: break 
frequency $\nu_b$, two Lorentzian low-frequencies $\nu_L$ and $\nu_{SL}$,
and two kilohertz QPO frequencies.  
%For example,
%on Figure~4, which illustrates the appearence of
%$\nu_b$,  $\nu_{SL}$ and  $\nu_{L}$, only one
%(higher) kilohertz QPO is seen. 
The presence of 
the second low frequency Lorentzian, which we refer to as $\nu_{SL}$,
has been reported in previous works on timing properties of 4U 1728-34 \citep[see][]{mig03,VS02,di01}.
In Figure \ref{ind_qpo} we present the observed  correlations
 of photon index $\Gamma$ vs $\nu_b$ (lower panel),  $\nu_L$
 (middle panel) and $\nu_{SL}$ (upper panel) for a  NS source 4U 1728-34.   
 The  index saturation seen in {\it BH sources}  at high values of $\nu_{L}$ (see TF04) is not observed in {\it this NS source}.
 The index seemingly increases without any sign of saturation as the low and break frequencies increase. 
One can naturally expect that in the high/soft state of an accreting NS source thermal 
equilibrium is achieved and that two blackbody components related to the disk and the NS would be 
seen in the data. 
 %(see e.g. Laurent \& Titarchuk 1999).
 Our data analysis confirms this expectation. In the NS high/soft state  the temperature 
of the Comptonizing cloud is  very close to the temperature of the NS blackbody component, 
namely a few keV. In this  case the blackbody soft photons do not gain any energy from the 
electrons of surrounding plasma cloud, i.e. the energy exchange between the electron gas to the 
photons is very small. The photon index of the Comptonization Green's function is
$\Gamma\gg1$ (see Sunyaev \& Titarchuk 1980, hereafter ST80). The blackbody spectra are 
only slightly Comptonized. Figure \ref{evol} shows the evolution of a spectrum when the 
mass accretion rate increases. When the source is in low/hard state the emerging radiation spectrum is 
formed as a result of Comptonization of soft photons generated in the disk
and neutron star atmospheres (see Figure \ref{soft_hard}, upper panel).
 In comparison with  low/hard BH spectra (for~which~$\Gamma\lax1.7$) the NS spectra are
slightly softer and have indices $\Gamma\gax 2$. 
%In  Figure \ref{soft_hard} (upper panel)
%we show the low/hard spectrum for 4U 1728-34 in blue.

In the NS case, which we analyze, the high/soft spectrum is the  
sum of two blackbody components, no power-law  
component is seen in the observed spectrum (see Figure \ref{soft_hard}, lower  
panel). It clearly indicates that the thermal equilibrium is
established when mass accretion rate is high.
 On the other hand in black holes the thermal equilibrium is
 not achieved in high/soft state because of the presence the event horizon in the system. 
%There is no condition for the equilibrium  
%between radiation and surrounding material, 
The emergent spectrum should deviate from the blackbody.
 \citet{grove} and BRT99 presented a set
of high/soft spectra for black holes. They showed that in addition
to the  soft (disk) blackbody component there is
an extended power-law component (for which $\Gamma\sim2.7$).  

%\subsection{Behavior of Fe K$_{\alpha}$ emission-line strength with the QPO frequencies}
 
%Matter goes in and nothing comes out and most of high energies photons comptonized in the bulk flow are trapped  
%there. As a result of this trapping along with the bulk Comptonization the specific power-law component is formed
%which photon spectral index saturates to the value of 2.5-2.9 (depending on the flow temperature) for high mass
%accretion rates $\dot M$ (see  Laurent \& Titarchuk 1999 and Turolla, Zane, \& Titarchuk 2003).   
%In BH sources the low and high frequency plateau regions of the
%data  are signatures of the two spectral phases which are explained by
%two different regimes of Comptonization ( upscattering): \ (1) bulk flow
%Comptonization in the soft state ( saturation at $\Gamma\sim 2.8$) and
%(2) thermal Comptonization in the hard state ( the photon index tends to level  at $\Gamma
%\sim1.7$). In  NS sources the spectral shape  and thus the index-frequency  is explained  
%by the thermal Comptonization only. In TF04 it was shown how the spectral index varies with the Compton cloud  
%optical depth $\tau_0$ (which is directly related to mass accretion rate) for NS.  
%It is easy to see from Fig. 1  that the observed behavior of the
%index as a function of $\nu_L$  and the index range are similar to the theoretical dependence of $\Gamma$
%vs $\tau_0$ in  TF04 (Fig. 5 there). One should keep in mind that $\tau_0$ and $\nu_L$ and consequently  
%$\dot M$ and $\nu_L$ are correlated with each other.

\section{On the nature of low-frequency oscillation modes}

In Figure 5 we present the observed ratio $\nu_{SL}/\nu_L$. It is apparent that this 
ratio  anticorrelates with $\nu_l$ and consequently with mass accretion rate.
 The number of points presented in Figure 5 is 118, for which the 
 correlation coefficient is  -0.52 and thus  the probability of non-correlation  of $\nu_{SL}$ and $\nu_L$   
is   $1.22\times10^{-8}$. The fit to the linear function gives us that $\nu_{SL}/\nu_L=(0.61\pm 0.01) -
(0.09\pm0.02)(\nu_L/100~{\rm Hz})$. The present analysis  indicates that 
the observed trend is statistically significant. The apparent trend in
 $\nu_{SL}-\nu_L$ relation does not fit within a simple harmonic-subharmonic picture.
%which indicates considerable anticorrelation.

We propose a model to explain this behavior of $\nu_L$ and $\nu_{SL}$ frequencies. We suggest 
that the accretion flow consists of two major geometrical components, namely, a 
%quasi-
spherical Compton corona
and a 
%quasi-
cylindrical (outflow) component related to the disk
(see Figure \ref{geometry}).
We treat  $\nu_L$ and its  $\nu_{SL}$  
as normal mode oscillation frequencies of spherical
and cylindrical components respectively. We assume
that spherical component is embedded into the cylindrical configuration and 
the corona radius  $R_{cor}$ is equal to the  cylinder radius  $R_{cyl}$, i.e. $R_{cor}\sim R_{cyl}=R_0$. 
The height of the cylinder  $H\gax R_0$.  

The wave equation for the displacement $u(t,{\bf r})$ \citep[e.g.][]{ll} reads
\begin{equation}
\frac{\partial^2 u}{\partial t^2}=a^2 \Delta u,
\end{equation}
where $a$ is sound speed in the plasma and ${\bf r}$ is a radius vector for a given point in the configuration.  
The Laplace operator, $\Delta$ has the form
\begin{equation}
\Delta=\frac{1}{r^2}\frac{\partial}{\partial r}\left(r^2\frac{\partial}{\partial r}\right)
\end{equation}
for the spherical geometry. In the cylindrical geometry $\Delta$ is  
\begin{equation}
\Delta=\frac{1}{r}\frac{\partial}{\partial r}\left(r\frac{\partial}{\partial r}\right)+\frac{\partial^2 }{\partial z^2},
\end{equation}
where $0<r<R_0$, and $0<z<H$. Two cases
of boundary conditions can be considered: $u(t,R_0)=0$ (fixed boundary), $\partial u/\partial r(t,R_0)=0$ (free boundary)
 for spherical geometry and  
$u(t,R_0,z)=u(t,r,H)=0$ (fixed boundary), $\partial u/\partial r(t,R_0,z)=\partial u/\partial z(t,r,H)=0$ (free boundary) 
for cylindrical geometry.

In order to find the eigen frequency for a given configuration  
one should solve the eigenvalue problem for Eq. (2) or (3) with the appropriate  
(fixed or free) boundary conditions. We  
are interested in the least (fundamental) eigen frequencies.
Using the separation variable method we present the solution of
Eq. (1) in the form $u(t,r)=T(t)X(r)$ for the spherical geometry,
where $X(r)$ can be found as a nontrivial (eigen) solution of  
equation
\begin{equation}
\label{sp_x}
\frac{1}{r^2}(r^2X')'+\lambda^2X=0
\end{equation}
for the homogeneous (fixed or free) boundary conditions. The solution of Eq. (\ref{sp_x}) is $X(r)=\sin \lambda r/r$ and 
the appropriate first (fundamental) eigenvalue is $\lambda_{fxs}=\pi/R_0$ for the fixed
 boundary condition $X(R_0)=0$. For the free boundary condition  
$X'(R_0)=0$ the eigenvalue $\lambda$ can be found as the first  
nontrivial root of transcendental equation  
$\tan \lambda R_0 = \lambda R_0$, which is  
$\lambda_{frs}=1.4303~\pi/R_0$.  

For the cylindrical geometry one can look for the solution as
 $u(t,r,z)=T(t)X(r)Z(z)$, where $Z(z)$ is a nontrivial solution
of equation  
\begin{equation}
Z''+\mu^2Z=0,
\end{equation}
and $X(r)$ is the solution of the following equation
\begin{equation}
\label{cyl_x}
\frac{1}{r}(r X')'+(\lambda^2+\mu^2)X=0,
\end{equation}
where $\mu^2$ and $\lambda^2+\mu^2$ are eigenvalues for
the eigenfunctions $Z(z)$ and $X(r)$ respectively.
With an assumption of symmetry of $Z(z)$
with respect to plane $z=0$ we obtain $Z(z)=\cos \mu z$  
 and for the fixed boundary  
condition we   have $Z(H)=\cos \mu_{fxc} H=0$  
which gives $\mu_{fxc}=\pi/2 H$.
 The bounded  
solution of Eq. (\ref{cyl_x}) in the interval
from $r=0$ to $r=R_0$  
is expressed through the Bessel function of zero order  
$X(r)=J_0 (\sqrt{\lambda^2+\mu^2}\, r)$. For the  
fixed boundary condition $\sqrt{\lambda_{fxc}^2+\mu_{fxc}^2}R_0=2.4$
 is the first root of equation  
$J_0 (\sqrt{\lambda_{fxc}^2+\mu_{fxc}^2}\, R_0)=0$. Then we obtain
 $\lambda$ as a function $R_0$ and $H$:
\begin{equation}
\label{lamb_fxc}
\lambda_{fxc}=\sqrt{(2.4/R_0)^2-(\pi/2H)^2}.
\end{equation}

For the free boundary condition we find $\mu_{frc}=\pi/H$ from the equation
$Z'|_{z=H}=[\cos (\mu_{frc} z])'|_{z=H}=0$. The eigenvalue  
$\lambda_{frc}^2+\mu_{frc}^2=(3.85/R_0)^2$ of Eq. (\ref{cyl_x})
can be found using the roots of equation $J'_0(\sqrt{\lambda_{frc}^2+\mu_{frc}^2}\, R_0)=J_1(\sqrt{\lambda_{frc}^2+\mu_{frc}^2}\, R_0)=0$, where we use
the recurrent relation for the Bessel functions (Abramovitz \& Stegun 1970,
formula 9.1.27). The expression for $\lambda_{frc}$ reads
\begin{equation}
\label{lamb_frc}
\lambda_{frc}=\sqrt{(3.85/R_0)^2-(\pi/H)^2}.
\end{equation}

The time-dependent component of the solution $u(t,{\bf r})$ is $T(t)=A \cos (2 \pi \nu t + \varphi_0 )$, where $\omega=2 \pi \nu=a \lambda$ is the fundamental 
rotational frequency related
to the fundamental eigenvalue $\lambda$. Evidently, the ratio of fundamental
frequencies  $\nu_c$ for cylindrical and $\nu_s$ for spherical
 configuration equals to the ratio of corresponding eigenvalues, i.e.  
$\nu_c/\nu_s=\lambda_c/\lambda_s$. For the fixed boundary conditions
the resulting ratio is
\begin{equation}
\label{ratio_c}
\nu_{fxc}/\nu_{fxs}=\lambda_{fxc}/\lambda_{fxs}=\sqrt{(2.4)^2-\pi^2(R_0/2H)^2}/\pi,
\end{equation}
while for the free boundary conditions we get
\begin{equation}
\label{ratio_s}
\nu_{frc}/\nu_{frs}=\lambda_{frc}/\lambda_{frs}=\sqrt{(3.85)^2-\pi^2(R_0/H)^2}/1.43\pi.
\end{equation}
%As we mentioned at the beginning of this section
Using Equations (\ref{ratio_c}) and (\ref{ratio_s}) in case of $H\gg R_0$  
one can obtain $\nu_{c}/\nu_{s}$ as $2.4/\pi=0.76$ and $3.85/1.43\pi=0.86$  
for the fixed and free boundary conditions respectively. For $R_0\sim H$
the corresponding ratios are 0.58 and 0.50.  

Our identification of the observed low frequency features $\nu_L$ and  
$\nu_{SL}$ as the fundamental oscillation modes of the spherical and cylindrical  
configurations sheds light on the nature of the observed dependence of 
$\nu_{SL}/\nu_L$ versus $\nu_L$ presented in Figure 5. In the low/hard  
state, when $\nu_L$ is about 10 Hz, the observed ratio reaches its highest values of about 0.8. 
The ratio steadily decreases from around 0.7 to values lower than 0.5 when the frequency
grows from 10 Hz to 120 Hz, i.e. when the source undergoes a  
transition from low/hard state to the high/soft state. Using the above
analysis of the behavior of $\nu_c/\nu_s$ as a function of
corresponding sizes of the spherical and cylindrical configurations ($R_0$ and $H$) 
one can argue that in the hard state the vertical size of the cylindrical configuration 
is much larger than the size of the spherical corona.
As the source progresses toward the soft state the  sizes of
configurations become comparable. We associate the cylindrical
configuration with the outflow and open field lines of the NS magnetosphere. The size of
this component should be related to the radio emission in the
source. We can thus conclude that the radio flux should be highest in the  
hard state and then to decrease steadily toward soft state.

\section{Disk-jet coupling in X-ray binaries and a link between Fe K$_{\alpha}$ emission-line strength and QPO frequencies}
The expected anticorrelation of X-ray spectral hardness and  
the radio brightness has  already been established by \citet{hom2004}
 for the low-mass X-ray binary GX 13+1 and by \citet{mirabel,klein}
 in the BHC GRS 1915+105. 
 %\citet{hom2004} a claim  that X-ray/radio pattern has 
%been found by Migliari et al. (2003) for 4U 1728-34  too. 
In fact, Homan et al. have observed the neutron star LMXB GX 13+1
simultaneously in X-rays and radio. The hard state has been linked with  radio fluxes, which were found to be 4 and 18 times higher than
the maximum detected in the soft state. This dependence of the radio flux on X-ray state has also been found  in 
4U 1728-34 by Migliari et al. (2003) who analysed 12 simultaneous radio and X-ray observations performed in two blocks in 2000 and 2001.
They  found that the strongest and most variable radio emission seems to be associated with transition between hard and soft states,
while weaker, persistent radio emission is observed when the source remains steadily in hard state.  In the hard state, when the break
frequency changes from 1 Hz to 10 Hz (see Fig. 2 and also Fig. 5), there are significant correlations between the radio flux density at
8.46 GHz and the 2-10 keV X-ray flux and also between radio flux density and the   break and low-frequency Lorentzian frequencies, $\nu_b$ and
$\nu_L$ in our definitions.   Migliari et al. also confirm previous findings that accreting neutron stars are a factor of $\sim 30$ less 
``radio loud'' than black holes.

Recently Miller  \& Homan (2005) claimed that the presence of Fe K$_{\alpha}$ emission-line strengths, QPOs, and the link between them
do not depend of jet activity (taking radio flux as a jet indicator) in GRS 1915+105.   We find that a similar  phenomenon takes
place in 4U 1728-34: while radio (and X-ray) emission strongly correlates with QPO frequencies (see Figs. 3, 4 in  Migliari et al.)
one can not see that Fe K$_{\alpha}$ equivalent width (EW) is correlated 
with QPO frequencies. 
In Figure 7 we present the behavior 
of the Fe K$_{\alpha}$ EW as a function of $\nu_b$.  Morever our EW range of 100-400 
eV is similar to that obtained 
in BHC GRS 1915+105. Miller \& Homan argue that the Fe K - connection has the potential to reveal the innermost regime in  an accreting
system. They also state that disk reflection models provide the best overall fit to the broadband spectrum in GRS 1915+105.
But it is not the case for 4U 1728-34. We find the spectra from this source are well fitted by 
a sum of two Comptonization components 
and the line component (see details in \S 2) and one does not need to include any additional 
components in the spectral model.
We find  that the intrinsic NS and disk emissions are completely scattered in the optically thick Compton 
cloud. The relative weight of the Comptonized component, $A/(1+A)$ 
is close to 1 as indicated by  our fitting to the data. On the other hand the Fe K$_{\alpha}$ EW is almost insensitive to QPO
values (taking QPO frequency value as a compactness indicator of X-ray continuum area).  
That leads one to conclude that the size of the Fe K$_{\alpha}$ emission area is weakly related 
to the compactness of X-ray continuum emission.  
%we can suggest that (at least) in 4U 1728-34 
This K$_{\alpha}$ emission  in 4U 1728-34 is possibly produced 
when  the emergent Comptonized spectrum illuminates  the wind (outflow) emanated from the binary.     
Laming \& Titarchuk (2004)  calculate the temperature and ionization balance in 
the outflow  from an accreting system under illumination by hard radiation from the central object. 
They  find that the equivalent width of Fe K$_{\alpha}$ line that originated in the wind is 
on the order of hundreds eV, close to
the observed values in 4U 1728-34. Laming \& Titarchuk also show that the strong iron line 
can be generated in the relatively cold extended region far away from the source of the illuminating photons 
(of order $10^3$-$10^4$ Schwarzchild radii).

\section{Conclusions}

We present a detailed spectral and timing analysis of
X-ray data for 4U 1728-34 collected with the {\it  RXTE}.
We find observational evidence for the correlation of
spectral index with low-frequency features: $\nu_b$, $\nu_{SL}$ and  
$\nu_{L}$. The photon index $\Gamma$ steadily increases from 1.8
in low/hard state to values exceeding 6.0 in the high/soft state.
Unlike in BH sources, there is no sign of saturation
 of the index in the high/soft state for 4U 1728-34. If this effect
was confirmed for other NS sources, it could be considered to be a neutron 
star signature. On the other hand, the presence of a photon-index saturation  (when the low QPO frequency increases) 
 would indicate the presence  a black hole. The two black-body component spectrum observed in the
 high/soft state can be also considered as a neutron star signature.

We also propose a model to explain the nature of the pair of  QPO low  
frequencies $\nu_L$ and  $\nu_{SL}$ and their behavior as
a function of the source spectral hardness. 
We argue that $\nu_L$  is a frequency of the radial oscillations of the quasi-spherical 
configuration around the NS  and $\nu_{SL}$  is a frequency of the radial and vertical 
oscillations of the cylindrical configuration there.
We show  that the values of  $\nu_L$ and  $\nu_{SL}$  
are determined by the sizes of the spherical  and cylindrical  
components of the accretion flow. Our analysis indicates that
the vertical size of the cylindrical component
is much larger than the corona radius in the low/hard state.
We suggest that the size of the cylindrical component is related  
to an outflow activity and, consequently, should be correlated with
 source radio flux. We expect the highest radio emission in  
the X-ray low/hard state.

While radio and X-ray emission and the X-ray spectral softness  strongly correlate with QPO frequencies, 
the Fe K$_{\alpha}$ line equivalent width does not show a significant correlation with
 QPO frequencies at least in NS binary 4U 1728-34.
It leads us to conclude that the compactness of the X-ray (radio) emission area 
(taking the QPO frequency value as a compactness indicator) is higher for
softer spectra (related to higher mass accretion rate). On the other hand the 
Fe K$_{\alpha}$ emission-line strength (EW) is almost insensitive to QPO
values (mass accretion rate) and one can conclude that the size of  
Fe K$_{\alpha}$ emission area is not related 
to the size of the X-ray continuum emission area. Thus the observations  
possibly suggest that the photospheric radius of the Fe K$_{\alpha}$ emission is
orders of magnitude larger than that for the X-ray continuum.

We appreciate productive discussions with Ralph Fiorito and Chris Shrader and we thank the referee for his/her constructive criticism and 
suggestions that improve  the paper presentation.

\newpage

\newpage

%\newpage
%\begin{figure}[ptbptbptb]
%\includegraphics[width=5.1in,height=5.1in,angle=0]{f1.eps}
%\caption{
%The observed  correlations
% of photon index vs break frequency $\nu_b$ (black), QPO low frequencies $\nu_L$
% (blue) and  $\nu_{SL}$ (red) in the NS source 4U 1728-34.    }
%\label{ind_qpo}
%\end{figure}
%\newpage

\newpage
\newpage
\begin{figure}[ptbptbptb]
\includegraphics[scale=0.65,angle=-90]{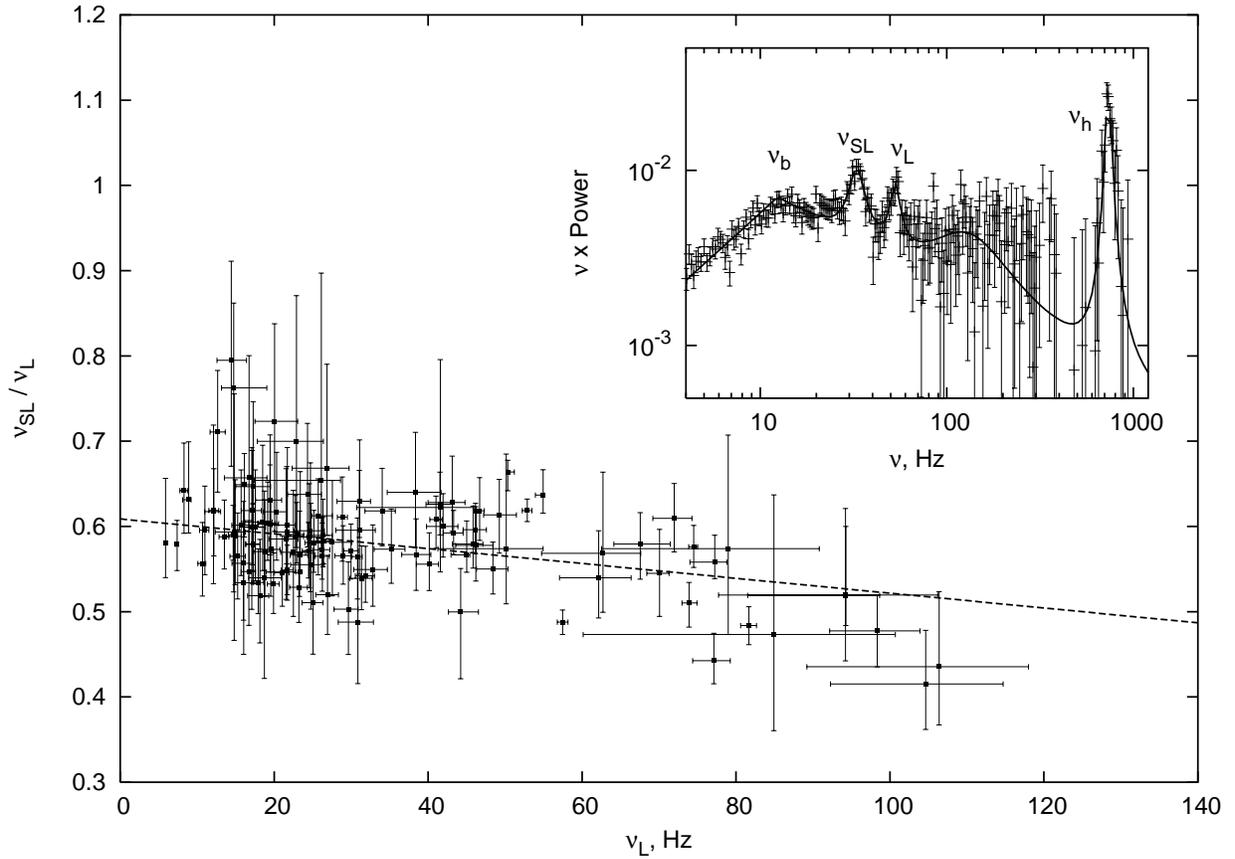}
\caption{
Observed ratio of $\nu_{SL}$  to $\nu_L$ versus of $\nu_L$.
Dashed lines indicates the linear fit to the data points. 
In the embedded panel a typical observed $\nu\times$power 
diagram with break and QPO frequency features is presented.}
\label{subhar}
\end{figure}

\newpage
\begin{figure}[ptbptbptb]
\includegraphics[scale=0.8,angle=0]{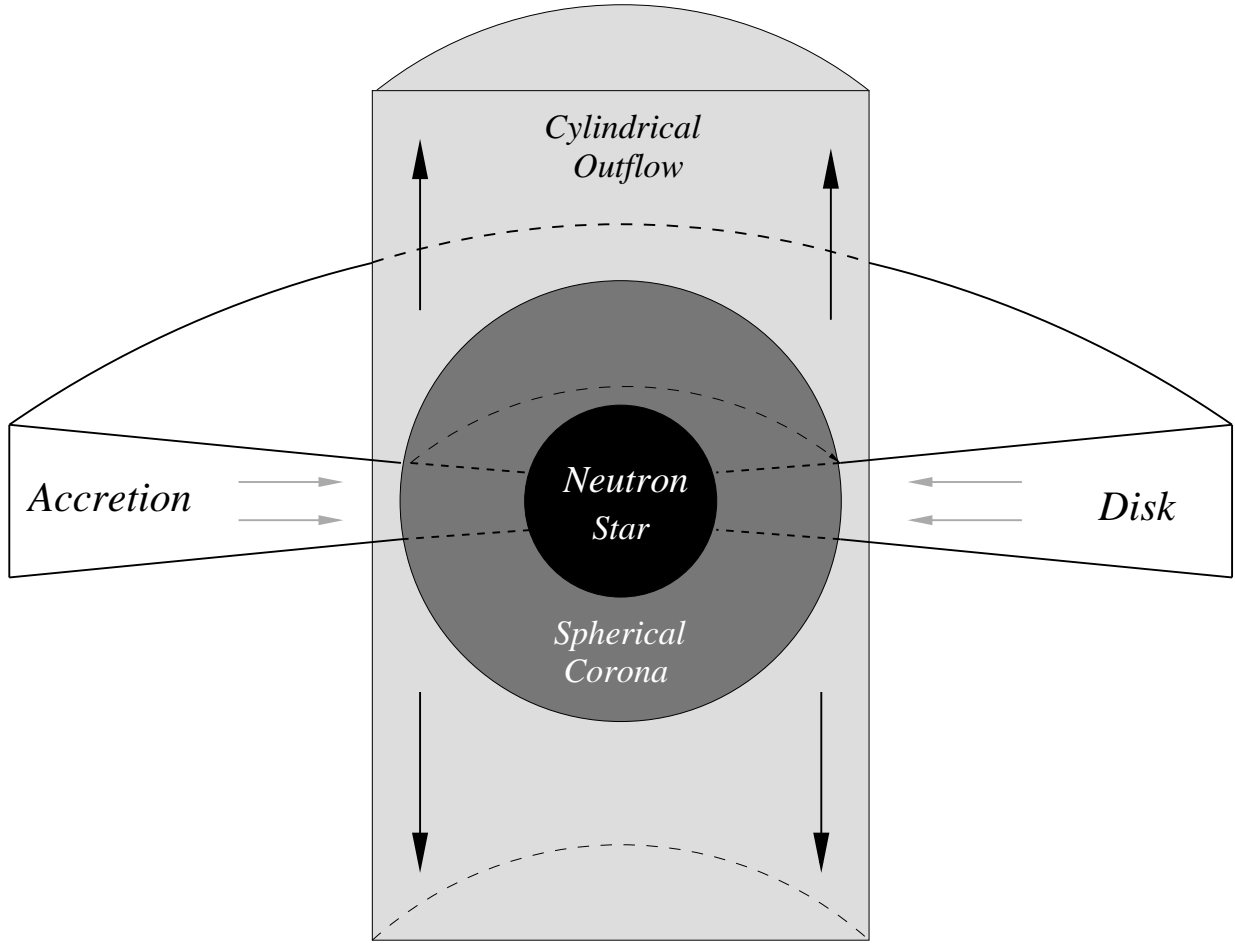}
\caption{
Graphical presentation of of NS-accretion flow system geometry. 
Two low-frequency oscillation modes related to $\nu_{L}$  to $\nu_{SL}$ are produced by radial vibrations
in spherical corona and  by radial and vertical vibrations in cylindrical outflow component respectively. 
\label{geometry}
}
\end{figure}

\newpage
\begin{figure}[ptbptbptb]
\includegraphics[scale=0.65,angle=-90]{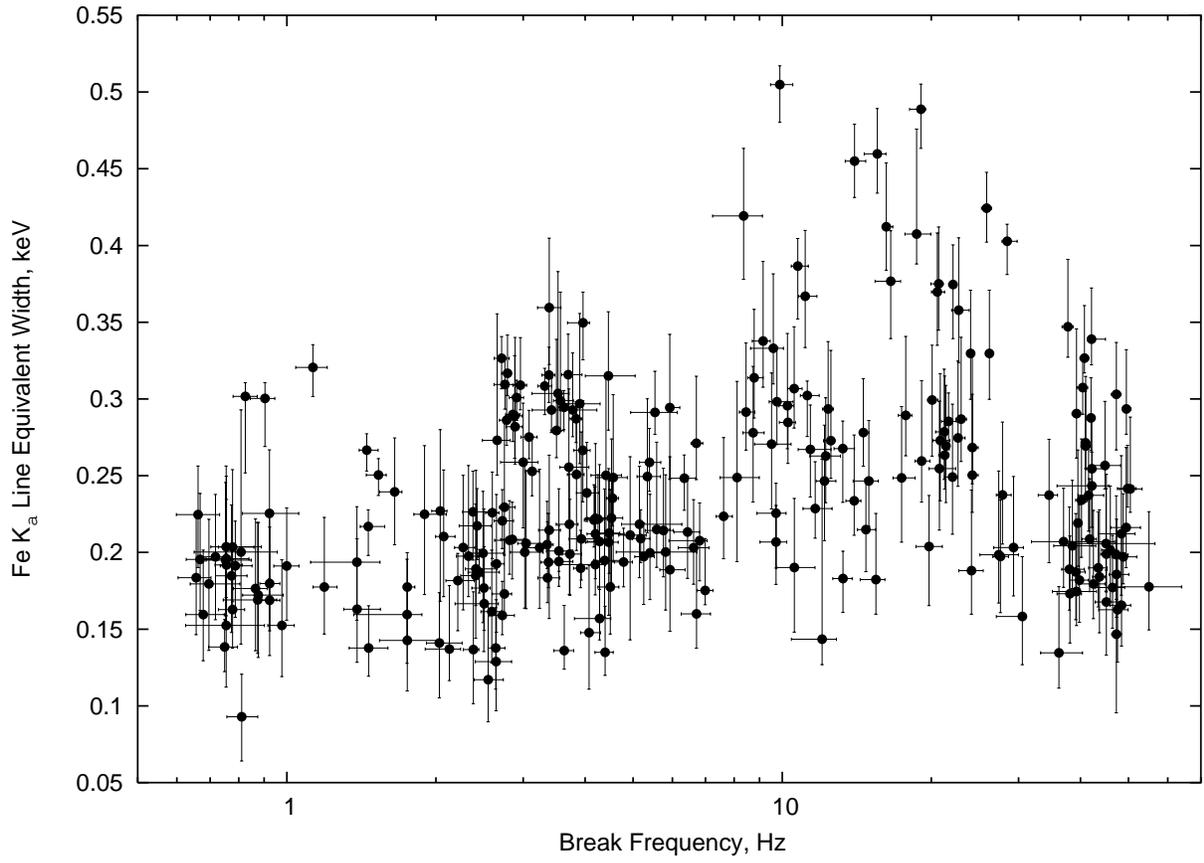}
\caption{
Observed  Fe K$_{\alpha}$ line equivalent width (EW) versus break frequency $\nu_b$.
}
\label{EWbreak}
\end{figure}

%\newpage
%\begin{figure}[ptbptbptb]
%\includegraphics[width=4.5in,height=5.5in,angle=-90]{f4.eps}
%\includegraphics[scale=0.65,angle=-90]{f4.eps}
%\caption{
%Observed ratio of  the low frequency $\nu_{SL}$  
%to  low frequency $\nu_L$ as a function of $\nu_L$.
%Two horizontal lines indicate the corridor where the most of ratio points are
%situated. In the embedded panel a typical observed $\nu\times$power diagram with break and QPO frequency features  is presented}
%\label{subhar}
%\end{figure}
\newpage
\begin{deluxetable}{lllrrrrl}
\tablewidth{0pt}
\tablecaption{Summary of RXTE archive data}
\tablehead{\colhead{Proposal ID} & \colhead{Start Date} & \colhead{Stop Date} & \colhead{Time, sec}& \colhead{$N_{obs}$}& \colhead{$N_{int}$}&\colhead{${\bar{N}_{PCUon}}$} &\colhead{refs.}}
\startdata
10073 & 15/02/96 & 01/03/96 & 253248 & 18 & 70 & 5.00 & 1,2,3,4,5,7\\
20083 & 19/09/97 & 01/10/97 & 193760 & 15 & 46 & 4.89 & 3,4,5,7 \\
30042 & 30/09/98 & 19/01/99 & 248352 & 31 & 68 & 5.00 & \\
40019 & 19/08/99 & 22/09/99 & 121472 & 16 & 33 & 3.31 & 6\\
40027 & 01/03/99 & 30/06/99 & 111904 & 17 & 32 & 4.27 & 6\\
40033 & 20/01/99 & 05/02/99 & 199008 & 19 & 42 & 4.51 & 6\\
50023 & 07/03/00 & 08/07/00 & 104336 & 32 & 34 & 3.98 & 6\\
50029 & 18/04/00 & 23/04/00 &  27168 & 6 & 9 & 4.05 & 6\\
50030 & 29/01/01 & 15/11/01 & 251088 & 29 & 65 & 3.75 & 6\\
\enddata
\tablerefs{(1) Strohmayer et al. 1996; (2) Ford \& van der Klis (1998); (3) van Straaten et al. (2002); (4) Di Salvo et al.
(2001); (5) Mendez, van der Klis \& Ford (2001); (6) Migliari, van der Klis \& Fender (2003); 
(7) Jonker, Mendez \& van der Klis (2000) }

%\citet{str96}, (2) \citet{fvk}, (3) \citet{VS02}, (4) \citet{di01}, (5) \citet{men01}, (6) \citet{mig03}, (7)\citet{jkr00}}
\end{deluxetable}
\end{document}